\documentclass[prl,aps,twocolumn,showpacs,superscriptaddress]{revtex4}  

\usepackage{graphicx}

\begin{document}

\title{One-Dimensional Transition Metal-Benzene Sandwich Polymers:\\
  Possible Ideal Conductors for Spin Transport} 

\author{H. J. Xiang}
\affiliation{Hefei National Laboratory for Physical Sciences at
  Microscale, 
  University of Science and Technology of
  China, Hefei, Anhui 230026, People's Republic of China}

\author{Jinlong Yang}
\thanks{Corresponding author. E-mail: jlyang@ustc.edu.cn}

\affiliation{Hefei National Laboratory for Physical Sciences at
  Microscale, 
  University of Science and Technology of
  China, Hefei, Anhui 230026, People's Republic of China}

\affiliation{USTC Shanghai Institute for Advanced Studies,
  University of Science and Technology of China,
  Shanghai 201315, People's Republic of China}

\author{J. G. Hou}
\affiliation{Hefei National Laboratory for Physical Sciences at
  Microscale, 
  University of Science and Technology of
  China, Hefei, Anhui 230026, People's Republic of China}

\author{Qingshi Zhu}
\affiliation{Hefei National Laboratory for Physical Sciences at
  Microscale, 
  University of Science and Technology of
  China, Hefei, Anhui 230026, People's Republic of China}
\affiliation{USTC Shanghai Institute for Advanced Studies,
  University of Science and Technology of China,
  Shanghai 201315, People's Republic of China}

\date{\today}

\begin{abstract}
  We investigate the electronic and magnetic properties of the proposed
  one-dimensional transition metal (TM$=$Sc, Ti, V, Cr, and Mn)
  -benzene (Bz) sandwich 
  polymers by means of density functional calculations. 
  [V(Bz)]$_{\infty}$ is found to be a quasi-half-metallic ferromagnet 
  and half-metallic ferromagnetism is predicted for [Mn(Bz)]$_{\infty}$.
  Moreover, we show that stretching the [TM(Bz)]$_{\infty}$ polymers
  could have dramatic effects on their electronic and  magnetic
  properties. The elongated [V(Bz)]$_{\infty}$ displays
  half-metallic behavior, and [Mn(Bz)]$_{\infty}$ stretched to
  a certain degree becomes an antiferromagnetic insulator. 
  The possibilities to stabilize the ferromagnetic order in [V(Bz)]$_{\infty}$ and
  [Mn(Bz)]$_{\infty}$ polymers at finite temperature are discussed.
  We suggest that the hexagonal bundles composed by these polymers
  might display intrachain ferromagnetic order at finite temperature by
  introducing interchain exchange coupling.
\end{abstract}


\maketitle

It is expected that the new generation of devices will exploit spin
dependent effects, what has been called spintronics
\cite{spintronics1,spintronics2}. 
A challenge now facing spintronics is transmitting spin signals
over long enough distances to allow for spin manipulation.
An ideal device for spin-polarized transport
should have several key ingredients. First, it should work well at
room temperature and should offer as high a magnetoresistance (MR)
ratio as possible.  
In this sense, a half-metallic (HM) ferromagnet with the Curie
temperature higher than room temperature is highly desirable since
there would be only one electronic spin channel at the Fermi
energy\cite{HM}.  
Second, the size or diameter of the materials should be
uniform for large scale applications.
Carbon nanotubes (CNTs) were considered as
promising one-dimensional (1D) spin mediators because of their
ballistic nature of conduction and relatively long spin scattering
length (at least 130 nm)\cite{MWCNT_exp,SWCNT_th}. Coherent spin
transport has been observed in 
multiwalled CNT systems with Co electrodes. The maximum
MR ratio of 9\%  was observed in multiwalled CNTs 
at 4.2 K. However, as the temperature increases to $20$
K, the MR ratio goes to zero, preventing any room temperature
applications\cite{MWCNT_exp}. 
A theoretical work suggested that the ferromagnetic (FM)
transition-metal (TM) /CNT hybrid structures may be used as devices
for spin-polarized transport to further increase the MR
ratio\cite{TM_CNT}. 
Unfortunately, although large spin polarization 
is found in these systems, there is no HM behavior.
Another difficulty with CNT is that the devices
are unlikely to be very reproducible due to the wide assortment of
tube size and helicity that is produced during synthesis. Wide
variation in device behavior was reported in the CNT
experiment\cite{MWCNT_exp}.

Recently, an experimental study 
suggested that the unpaired electrons on the
metal atoms couple ferromagnetically in the multidecker
organometallic sandwich V-benzene (Bz) complexes, i.e.,  V$_n$(Bz)$_{n+1}$ clusters\cite{exp1}. The
FM sandwich clusters are supposed to serve as
nanomagnetic building blocks in applications such as 
recording media or spintronic devices. 
A subsequent density functional study confirmed the FM
coupling in multidecker sandwich V$_n$(Bz)$_{n+1}$ clusters\cite{th1}. 
Motivated by the above experimental and theoretical studies,
we propose that the 1D organometallic sandwich polymers
[V(Bz)]$_{\infty}$ are the possible candidates for spintronic devices for
spin-polarized transport since the polymers with an inherently uniform
size are supposed to be also FM.

Metal-ligand molecules have been the subject of many studies in the
past decade\cite{huckle,Rao_Ni,Bz1,exp2,TM_Bz,Mn_Bz,JCP_V,La_Bz}.  
Especially, TM-Bz 
complexes (M$_{n}$(Bz)$_{m}$) are the prototypical
organometallic complexes for studying the d$-\pi$ bonding interactions. 
Depending on the metal, there are two types of structures for
M$_{n}$(Bz)$_{m}$: Multiple-decker sandwich structures and
metal clusters  fully covered with Bz molecules (rice-ball
structures). The former sandwich structure is characteristic of the
complexes for early transition metals (Sc-V), whereas the latter is
formed for late transition metals (Fe-Ni). Cr(Bz)$_2$ and
Mn(Bz)$_2$ with a sandwich structure are also observed 
experimentally\cite{exp2}.
Although the multidecker sandwich M$_{n}$(Bz)$_{m}$ clusters have been
extensively studied theoretically\cite{TM_Bz,Mn_Bz,JCP_V,La_Bz},
however, studies of the 1D [TM(Bz)]$_{\infty}$ polymers are
still very lacking.
To our knowledge, only one semi-empirical H\"{u}ckle calculation
was performed to examine the thermodynamical stability of the 1D
[TM(Bz)]$_{\infty}$ polymers\cite{huckle}. The electronic and
magnetic properties of the 1D [TM(Bz)]$_{\infty}$ polymers
remain to be explored using  
{\it{ab initio}} quantum mechanics methods. Moreover, the differences
in bonding, energetics, and magnetic properties
between [TM(Bz)]$_{\infty}$ polymers with different TM atoms are to
be clarified since the properties of TM-Bz sandwich clusters 
depend on the number of the 3d electrons of the TM atoms. 

In this paper, we perform a comprehensive first
principles study on the electronic and magnetic properties 
for the
proposed 1D sandwich polymers. Only the [TM(Bz)]$_{\infty}$ polymers 
with TM $=$ Sc, Ti, V, Cr, and Mn are considered 
since Fe, Co, and Ni 
usually react with Bz to form the rice-ball structures\cite{Rao_Ni,exp2}.   
Our theoretical calculations are performed within spin-polarized
density-functional theory (DFT) with the
generalized gradient approximation (GGA) PW91 functional\cite{pw91}.
The Vienna {\it {ab initio}} simulation package (VASP) \cite{vasp1,vasp2},
a plane-wave based program, is used.
We describe the
interaction between ions and electrons
using the  frozen-core projector augmented wave approach
\cite{paw}. 
The plane-wave basis set cut off is $400$ eV.
In a typical calculation, a 1D periodic boundary condition is
applied along the polymer axis (the $z$ direction) with Monkhorst-Pack\cite{mp} k-point
sampling. 

\begin{figure}[!hbp]
  \includegraphics[width=8.5cm]{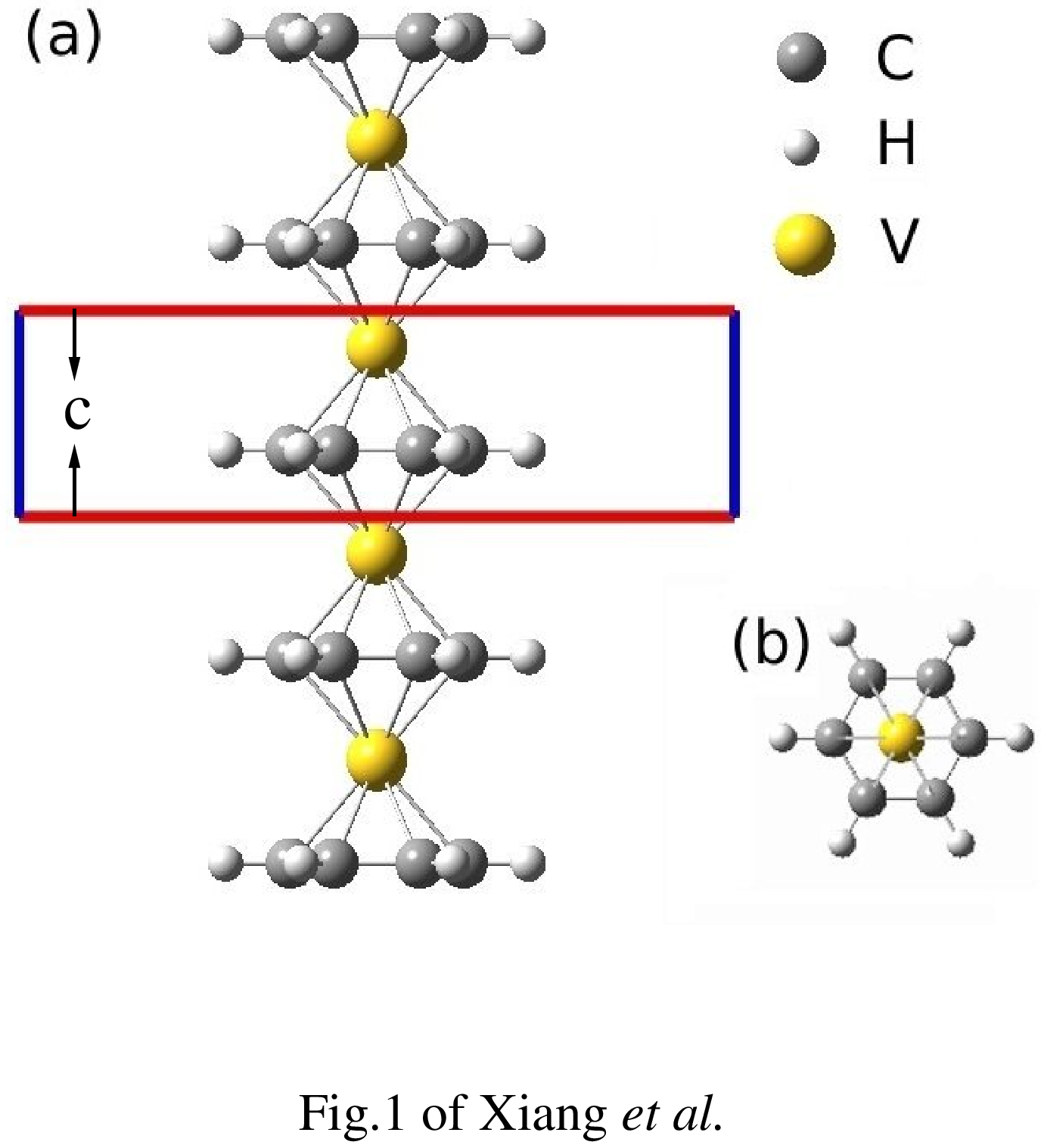}
  \caption{(Color online) Structure of the D6H [V(Bz)]$_{\infty}$ sandwich
    polymer (a) side view and (b) top view. The enclosed region in
    (a) indicates the unit cell of the polymer and $c$ denotes the
    lattice constant.}
  \label{fig1}
\end{figure}

There are two typical structural configurations for the sandwich
polymers.
One is a normal sandwich structure with D6h
symmetry, in which the TM atom and Bz rings are arranged
alternatively, as shown in Fig.~\ref{fig1} for the the D6h
[V(Bz)]$_{\infty}$  polymer.
The other is a staggered sandwich structure
(D6d symmetry) in which one of the Bz rings is rotated
by 30\% with respect to the other ring.
Our preliminary calculations for both D6h and D6d [V(Bz)]$_{\infty}$
polymer indicate that the electronic and magnetic properties of the D6d
conformer differ little from those of the corresponding D6h one.
The qualitative similarity between the D6d and D6h configurations is
also found in previous study on the sandwich clusters\cite{JCP_V}.
So hereafter we mainly focus on the polymers with D6h symmetry.

Our results for different [TM(Bz)]$_{\infty}$ polymers are
summarized in Table~\ref{table1}. 
We can see that the sandwich [TM(Bz)]$_{\infty}$ polymers display
rich electronic and magnetic properties.  
All [TM(Bz)]$_{\infty}$ polymers studied here are metallic except
that [Cr(Bz)]$_{\infty}$ is a non-magnetic(NM) insulator with a $0.71$ eV
direct energy gap at $\Gamma$.
[Sc(Bz)]$_{\infty}$ is paramagnetic (PM).  
For [Ti(Bz)]$_{\infty}$, the AFM state is slightly favorable by $6$
meV/Ti over the FM state.
The ground state for
[V(Bz)]$_{\infty}$ and [Mn(Bz)]$_{\infty}$ is the robust FM state.
The diversity of the magnetic properties for the sandwich clusters
with different TM atoms has been found by previous theoretical
studies.
The NM insulating property for [Cr(Bz)]$_{\infty}$ accords
with the singlet state for Cr(Bz)$_2$ due to the 18-electron rule
\cite{TM_Bz}.
For [V(Bz)]$_{\infty}$ and [Mn(Bz)]$_{\infty}$, the FM ground state
also agrees qualitatively with the doublet ground state for V(Bz)$_2$ and
Mn(Bz)$_2$\cite{Mn_Bz, TM_Bz}. 
However, two distinct cases are Sc and Ti.
Previous study showed that Sc(Bz)$_2$ is a doublet state in contrast
with the PM state for [Sc(Bz)]$_{\infty}$ found here.
On the other hand, Ti(Bz)$_2$ is a singlet state contrasting shaply 
to the AFM ground state for [Ti(Bz)]$_{\infty}$. 
The binding energy  of the [TM(Bz)]$_{\infty}$ polymers as given
in Table~\ref{table1} is defined as: 
$E_{b}=E$(Bz)$+E$(TM) - $E$([TM(Bz)]$_{\infty}$), where $E$(TM) is the
energy of the isolated TM atom. We can see that
[V(Bz)]$_{\infty}$ has large thermodynamic stability with the largest
binding energy ($5.334$ eV/V atom). 
The trend of the lattice constant for the [TM(Bz)]$_{\infty}$
polymers is similar with that of the distance of the TM atoms from
the center of the Bz ring in TM(Bz)$_2$\cite{TM_Bz}. 
[Cr(Bz)]$_{\infty}$ has the the smallest lattice constant ($3.30$
\AA) but not very large binding energy, which might result from
the stable 3d$^{5}$4s$^{1}$ valence configuration for the Cr atom.

\begin{figure}[!hbp]
  \includegraphics[width=8.5cm]{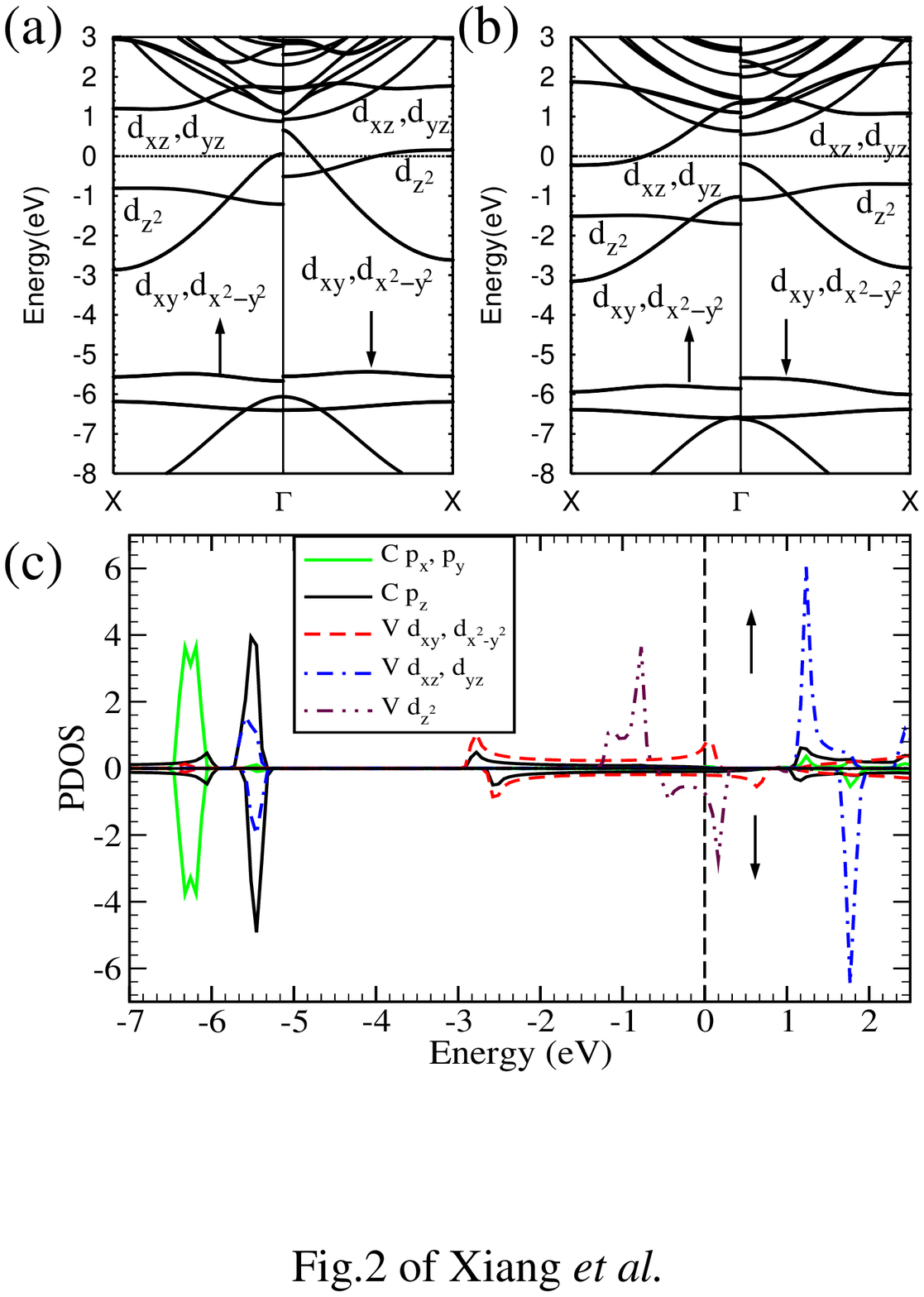}
  \caption{(Color online) Electronic band structures for (a) [V(Bz)]$_{\infty}$
    and (b) [Mn(Bz)]$_{\infty}$. (c) shows the C and V PDOS
    for [V(Bz)]$_{\infty}$.}
  \label{fig2}
\end{figure}

Since [V(Bz)]$_{\infty}$ and [Mn(Bz)]$_{\infty}$ have robust ferromagnetism,
which is crucial for practical spintronic applications, the
detailed electronic and magnetic properties are examined carefully. 
The band structures for [V(Bz)]$_{\infty}$ and [Mn(Bz)]$_{\infty}$
are shown in Fig.~\ref{fig2}. 
For [V(Bz)]$_{\infty}$, there is a gap just
above the Fermi level for the spin-up component indicating a quasi-HM
behaviour.
The V, Mn 4s orbitals lie in the conduction band with very
high energy and the bands around the Fermi level are mainly
contributed by the TM 3d orbitals. 
The bands which are mainly composed by the TM 3d orbitals are
labeled by their 3d components in Fig.~\ref{fig2}.
In the crystal field of the Bz ligands, the TM 3d
bands split into three parts: A two-fold degenerated band D1
contributed mainly by TM 3d$_{xy}$ and 3d$_{x^{2}-y^{2}}$ orbitals,
a non-bonding band D2 with TM 3d$_{z^{2}}$ character, and a
two-fold degenerated anti-bonding band D3 due to TM 3d$_{xz}$ and
3d$_{yz}$.
The dispersive D1 band crosses the localized D2 band. 
And the D3 band with high energy is separated from the other two
bands. 
For [V(Bz)]$_{\infty}$, in the majority part, the D2 band is fully
occupied and there is a small hole in the D1 band, and 
both D1 and D2 bands are partly filled in the minority part.
Such 3d orbitals occupation results in $0.80$ $\mu_{B}$ total magnetic
moment for [V(Bz)]$_{\infty}$.
Since Mn atom has two valence electrons more than V atom,
both D1 and D2 bands are fully occupied in
both spin components in [Mn(Bz)]$_{\infty}$. The remained one
valence electron half fills the D3 band in the spin majority part. 
Such a picture of the orbital occupation leads to the peculiar HM FM
behavior in [Mn(Bz)]$_{\infty}$ with a total magnetic moment $1.00$
$\mu_{B}$. 

The above analysis based on the crystal field theory
qualitatively describes the electronic and magnetic properties of the two
polymers. However, the large dispersion of the D1 band and the FM
mechanism in these systems remain to be clarified. To penetrate into
this problem, we 
plot the projected density of states (PDOS) for [V(Bz)]$_{\infty}$ in
Fig.~\ref{fig2}(c).
We can see there is a considerable contribution to the V D1 band
near $X$ from C 2p$_{z}$ orbitals, however at $\Gamma$ the V
3d$_{xy}$ and 3d$_{x^{2}-y^{2}}$ orbitals could only hybrid with C
2p$_{x}$ and 2p$_{y}$ orbitals due to
the symmetry matching rule. 
The C 2p$_{z}$ orbitals of Bz hybrid with  V 3d$_{xy}$ and 3d$_{x^{2}-y^{2}}$
orbitals to result in bonding states near $X$, and anti-bonding like
states near $\Gamma$ arise due to the 
weak hybridization between Bz 2p$_{x}$ and 2p$_{y}$ states and these V 3d
orbitals. Such hybridization leads to the large dispersion of the D1
band with the highest energy at $\Gamma$ in the whole
Brillouin zone.  
Clearly, the contribution from Bz orbitals to the states near the Fermi
level is very small, and the holes are mainly of TM 3d character,
so ferromagnetism in such systems is identified to be due to the
double exchange (DE)
mechanism\cite{DE,DE1}.

\begin{figure}[!hbp]
  \includegraphics[width=8.5cm]{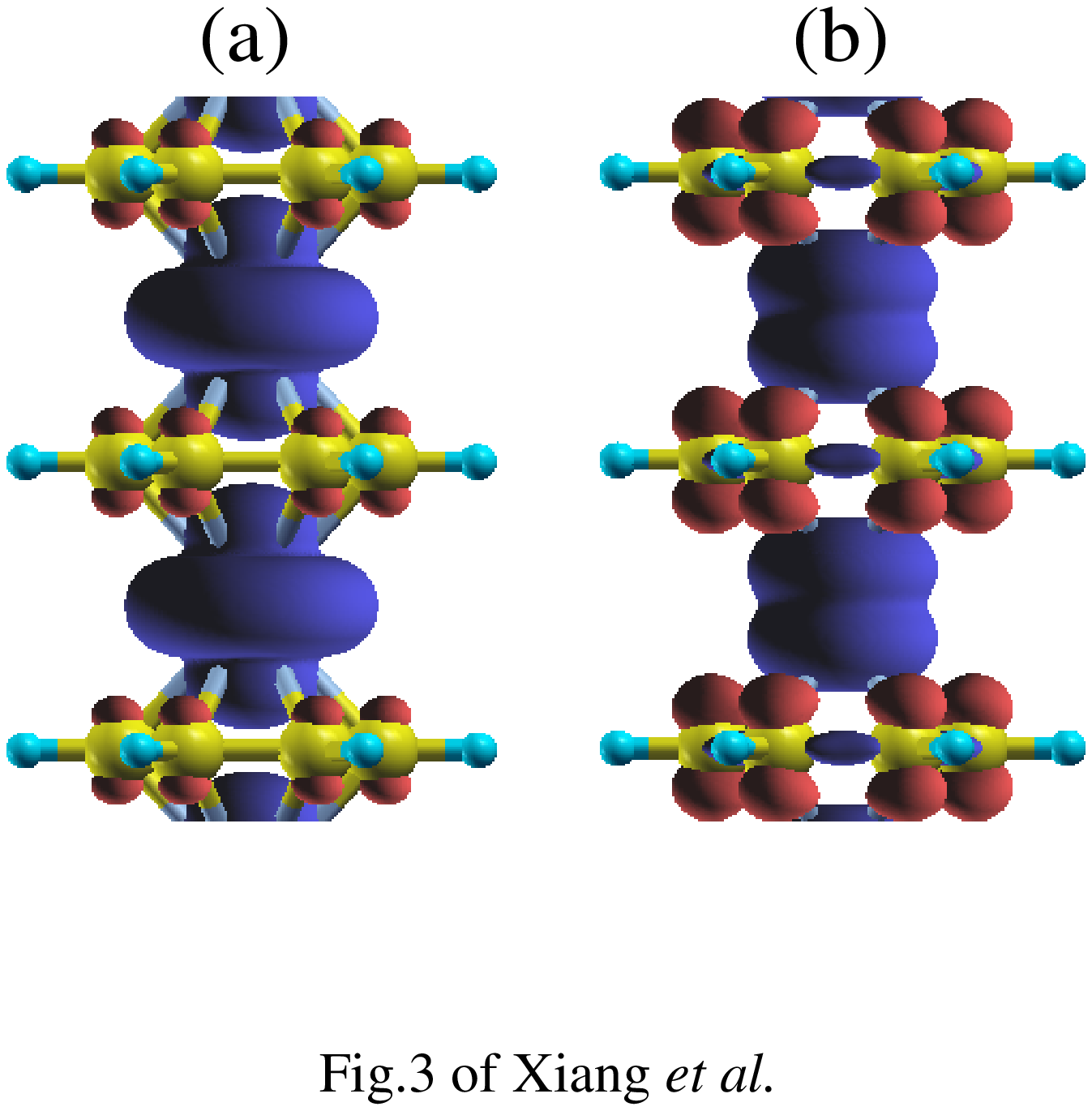}
  \caption{(Color online) Spin density for (a) [V(Bz)]$_{\infty}$ and (b)
    [Mn(Bz)]$_{\infty}$. The isovalue for the red and blue
    isosurfaces is $-0.014$ and $0.014$ $e$/\AA$^{3}$ respectively.}
  \label{fig3}
\end{figure}

The spin densities for both [V(Bz)]$_{\infty}$ and [Mn(Bz)]$_{\infty}$
are shown in Figs.~\ref{fig3} (a) and (b) respectively. The spin
density for [V(Bz)]$_{\infty}$ clearly indicate the 3d$_{z^{2}}$
character, which is accord with the band structure. Differently, the
spin density for [Mn(Bz)]$_{\infty}$ is mainly due to Mn 3d$_{xz}$ and
3d$_{yz}$ orbitals, as discussed above. An interesting phenomenon
should be noted: The spin polarization not only locates around the TM
atoms, but also has a small negative contribution from the Bz
2p$_{z}$ orbitals. The spin injection to Bz is due to the
hybridization effect. More occupied hybridized bands will lead
to larger spin polarization of Bz. 
Since the D3 band of the spin up component is also half filled, 
the spin polarization of Bz in [Mn(Bz)]$_{\infty}$ is (0.090
$\mu_{B}$) larger than  that (0.067 $\mu_{B}$) in [V(Bz)]$_{\infty}$.
The spin injection to Bz has also been found by 
Muhida {\it et al.} in the study of TM(Bz)$_2$ (TM=Mn, Fe, and
Co)\cite{Mn_Bz}. However,
the six C atoms in a Bz ring contribute
equally to the spin polarization in both [V(Bz)]$_{\infty}$ and
[Mn(Bz)]$_{\infty}$, contrasting sharply to their
results that two of the C atoms contribute much larger than the
others\cite{Mn_Bz}.     

In the above paragraphs, we have shown that the ground state of both
[V(Bz)]$_{\infty}$ and [Mn(Bz)]$_{\infty}$ polymers is FM.
We now turn to study the magnetic properties of [V(Bz)]$_{\infty}$ and
[Mn(Bz)]$_{\infty}$ polymers at finite temperature.
Unfortunately, there is no long-range FM or AFM order in
any infinite strictly 1D isotropic systems with finite range exchange
interaction at any nonzero temperatures\cite{Mermin}.
However, there are several mechanisms for
the occurrence of FM or AFM order in 1D or quasi-1D
systems at finite temperature: barriers due to the magnetic anisotropy, finite size, and
interchain couplings. Large
magnetic anisotropy energy can block the thermal fluctuation and
stabilize long-range FM order,
e.g., 1D
monatomic chains of Co constructed on a Pt substrate exhibit
long-range FM
order below the blocking temperature 15 K\cite{Co}.
However,
our calculated results show that the magnetic anisotropy energy in
both [V(Bz)]$_{\infty}$ and
[Mn(Bz)]$_{\infty}$ polymers
is smaller than 0.1 meV/TM, indicating that the ferromagnetism in these polymers
might be stabilized only at extremely low temperature by the magnetic
anisotropy induced barrier.
There are also some reports, indicating the
existence of FM in 1D chains with
finite length, where the length of the chains is related to the range
of the exchange interaction\cite{Co}. The most relevant work related to the current study is
the discovery of FM in 1D V$_n$(Bz)$_{n+1}$ 
sandwich clusters by Miyajima {\it et al.} \cite{exp1} as mentioned above.
Since here we
mainly focus on infinite 1D polymers, thus we won't discuss more on
this point.
The interchain coupling is found to be an important mechanism to
stabilize long-range FM or AFM order in quasi-1D systems.
Some quasi-1D systems have been discovered
to display FM \cite{FM_1D_1,FM_1D_2} or AFM \cite{AFM_1D}
order at finite temperature. Here we study the hexagonal bundles composed by
[V(Bz)]$_{\infty}$ or
[Mn(Bz)]$_{\infty}$ polymers.
The optimized distance between the
centers of neighour polymers is about 6.95 \AA\ for both V-Bz and Mn-Bz
bundles.
Our results indicate that these bundles are energetically more favorable (by
about 31 and 17 meV for Mn-Bz and V-Bz  bundles respectively) than their
corresponding 1D isolated chains. 
It is well known that there is spin frustration for a AFM spin
configuration in a hexagonal lattice.
Here we refer to the spin configuration where some neighbour chains
couple antiferromagnetically and some neighbour chains
couple ferromagnetically as the AFM one.
The AFM coupling 
between V-Bz chains is slightly 
more favorable by about 3 meV/TM than the FM coupling. In contrast,
the interchain coupling in 
Mn-Bz bundles is found to
be FM (with energy about 5 meV/TM lower than the AFM coupling). 
The different coupling manner
between these two cases might originate from the different characters
of their spin densities. 
From the energy difference between the AFM and FM coupling in these
bundles, the interchain coupling $J$ 
is estimated to be about 0.75 or 1.25 meV/TM for V-Bz and Mn-Bz bundles respectively.
The interchain coupling should be large enough to stabilize the
 intrachain FM at finite 
 temperature since it  is 
 significantly larger than that in other typical qausi-1D magnetic
 systems, for example, the interchain
 coupling $J$ was estimated to be about 0.1 K (0.009
 meV) in $p$-nitrophenyl nitronyl nitroxide ($p$-NPNN) with a
 Curie temperature 0.65 K\cite{FM_1D_2}.

The intrachain exchange coupling is an important parameter in
describing the magnetic properties in quasi-1D
magnetic systems.
Here we derive the exchange parameters in
[V(Bz)]$_{\infty}$ and [Mn(Bz)]$_{\infty}$ isolated polymers using
total energy calculations for different spin configurations.
The intrachain coupling in their bundles is expected to have similar 
magnitude as that in isolated polymers.
Since the distance between a TM atom and its next nearest
neighbour TM is larger than $6.0$ \AA, the magnetic coupling
between them should be negligible.
Considerating only the nearest magnetic coupling, the effective
Heisenberg Hamiltonian for the 1D [TM(Bz)]$_{\infty}$ polymer can be
written as  $H_{eff}=- \sum_{i} J \mathbf{e}_{i} \cdot \mathbf{e}_{i+1}$,
where $J$ is the effective exchange parameter between the nearest
neighbour TM atoms,
and $\mathbf{e}_{i}$ is the unit vector pointing in the direction of
the magnetic moments at site $i$. Here we term $J$ as an
effective exchange parameter, but not a pure exchange parameter
since $J$ also includes the magnitude of the magnetic moment of
the TM atoms.    
One might expect that $J$ could be deduced from the
energy difference between the AFM state and FM state.
However, we find that the local magnetic moments for V and Mn atoms in
the AFM state differ significantly from those in the FM state, so
the above method to get $J$ might be questionable.
We circumvent this problem by considering another magnetic
configuration: The magnetic configuration is $\uparrow \uparrow
\downarrow \downarrow$ with the periodicity four times of that for
the FM case. In this case, the local magnetic moments for V and Mn
atoms are almost the same as those in the FM state. 
The energy difference between this configuration and
the FM state is $J$ per TM atom.
The estimated effective exchange parameter for [V(Bz)]$_{\infty}$ and
[Mn(Bz)]$_{\infty}$ is 24 meV and 62 meV respectively, which is larger
than the intrachain exchange coupling (about 0.4 meV) in $p$-NPNN\cite{FM_1D_2}.
The relatively large effective exchange parameter for [V(Bz)]$_{\infty}$ and
[Mn(Bz)]$_{\infty}$ is due to the DE mechanism, which
usually lead to FM with a relatively high Curie temperature,
e.g. the relative high Curie temperature (about $228$-$370$ K)\cite{GaMnN}
for Mn doped GaN is due to the DE mechanism\cite{DE1,GaMnN_th}.

\begin{figure}[!hbp]
  \includegraphics[width=8.5cm]{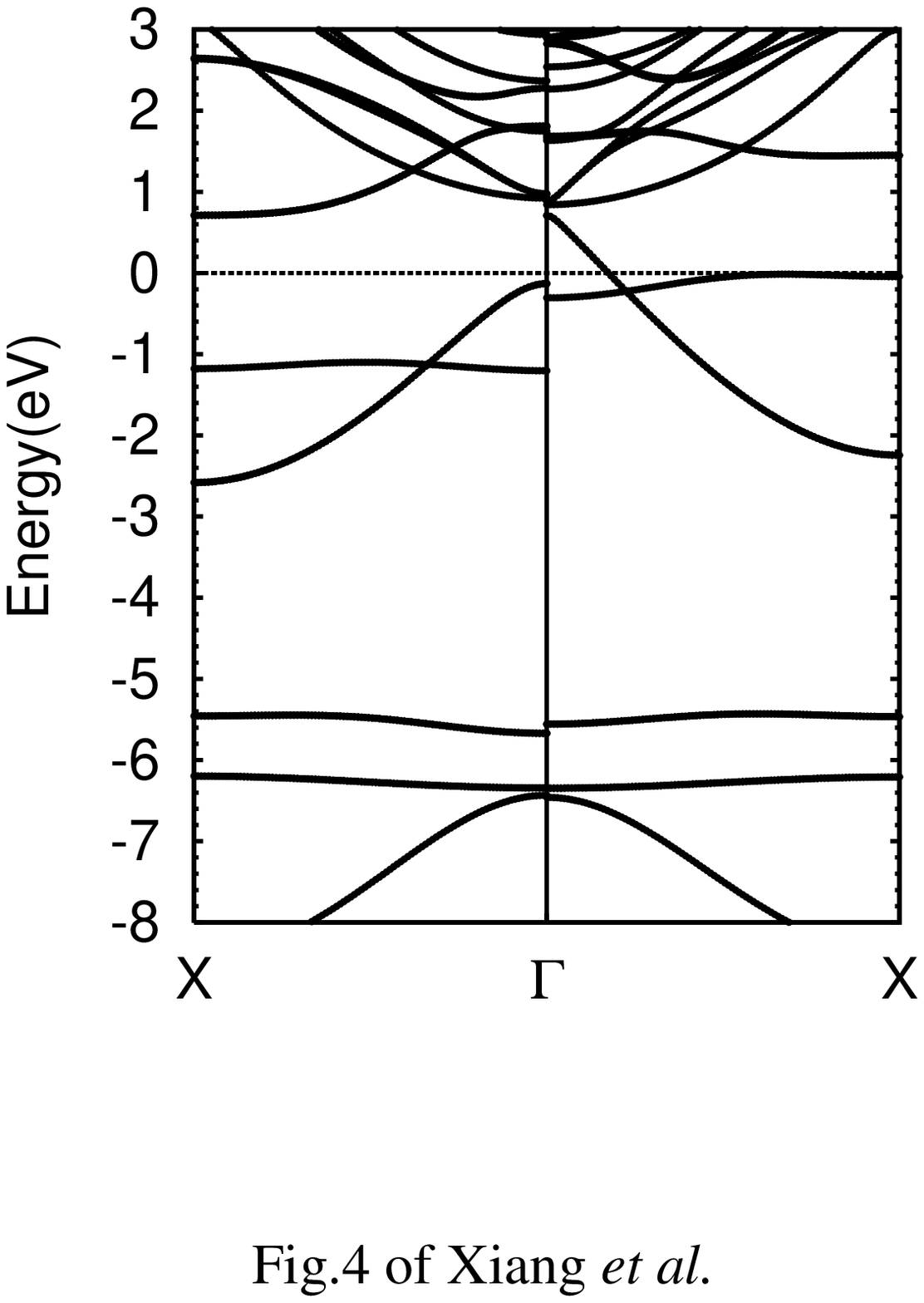}
  \caption{Electronic band structure for [V(Bz)]$_{\infty}$ with
    the lattice constant $c=3.55$ \AA. It clearly shows the HM
    behavior in this system.}
  \label{fig4}
\end{figure}

Pressure is well known to have drastic effect on the electronic and magnetic properties 
of the magnetic materials\cite{pres}. 
Here, the effect of pressing and stretching on the electronic and magnetic properties 
of [V(Bz)]$_{\infty}$ and
[Mn(Bz)]$_{\infty}$ is examined by changing the lattice constant. 
It is found that the electronic and magnetic properties of [V(Bz)]$_{\infty}$ and [Mn(Bz)]$_{\infty}$ 
change little when they are compressed to a reasonable degree.
However, stretching can dramatically change their electronic and magnetic properties.
[V(Bz)]$_{\infty}$ turns into HM FM when the lattice
constant is in the range $3.47$ to $3.78$ \AA.
To show the HM FM behavior, we plot 
the band structure of [V(Bz)]$_{\infty}$ with a lattice constant
3.55 \AA\ in Fig.~\ref{fig4}. Clearly, the energy gap opens in the
spin majority part in the stretched [V(Bz)]$_{\infty}$ polymer,
in contrast to a gap opening in the spin minority part in
[Mn(Bz)]$_{\infty}$.  
Along with the increase of the lattice constant, the hybridization
between TM and Bz is weaken. The band width of the D1 band,
especially in the spin majority part, becomes narrower. In the spin
majority part the downshift of the D1 band near $\Gamma$ 
makes the D1 band fully occupied and the system HM ferromagnetism.
We note that turning  [V(Bz)]$_{\infty}$ polymer into HM FM through
stretching experimentally the polymer is possible.
The stress in a hexagonal
bundle composed by [V(Bz)]$_{\infty}$ polymers with lattice constants
$a=6.95$ \AA\  and $c=3.55$ \AA\ is about $4.8$ GPa, which could be easily
realized experimentally.    
Increasing the lattice constant of [Mn(Bz)]$_{\infty}$ to a critical
value $3.48$ \AA, the HM behavior disappears. Further increasing the
lattice constant of [Mn(Bz)]$_{\infty}$ up to a second critical
value $3.70$ \AA, a FM metal to AFM insulator transition takes place.
The drastic changes of the  electronic and magnetic properties
of [Mn(Bz)]$_{\infty}$ is also caused by the
reduced hybridization. Although the band narrowing of the D1 band
also takes place, however, the downshift of the D3 band near $X$ in
the spin majority part plays a crucial role in determining the
electronic and magnetic properties of [Mn(Bz)]$_{\infty}$.  
The disappearance of the HM behavior results from the partly
occupying of the D1 band in the spin minority part as a consequence
of the downshift of the D3 band in the spin majority part.
The FM metal to AFM insulator transition can be reasonably explained
as: The band narrowing of the D1 band results in smaller DE effect,
thus the superexchange AFM interaction can compete  
with and even overwhelm the DE interaction since the
former depends on hopping linearly, and the latter quadratically.

Previous studies indicated that the local density approximation (LDA) and other semilocal GGA
functionals fail to describe some conducting polymers. 
A typical example is the dimerization of trans-polyacetylene: LDA
predicts almost no carbon-carbon bond length alternation (BLA),
whereas Hartree-Fock (HF) produces a too large
BLA\cite{polyacetylene}. It was shown that 
hybrid density functional mixed with
exact exchange, namely B3LYP\cite{B3LYP}, provided
dimerization parameters close to experiment\cite{polyacetylene}. 
Thus it is interesting to know the electronic and magnetic properties of the polymers within the B3LYP 
formalism. For this purpose, we perform some calculations on
these 1D polymers using CRYSTAL03
code\cite{CRYSTAL03}.
The results show that both [V(Bz)]$_{\infty}$ and [Mn(Bz)]$_{\infty}$
are HM FM
in their most stable configurations. Although there are some minor
differences between the PW91 and B3LYP results, our main conclusions
remain valid when using the B3LYP functional.
The applicablity of the semilocal GGA functional to the 1D
[TM(Bz)]$_{\infty}$ polymers is not surprising since
the current studied systems are not $\pi$-conjugated. In fact, LDA has been 
successfully applied to study the conformation and electronic structure of
polyethylene\cite{polyethylene}.

In conclusion, we have performed a comprehensive first
principles study on the electronic and magnetic properties for the
proposed 1D [TM(Bz)]$_{\infty}$ polymers with TM= Sc, Ti, V, Cr, and
Mn. We demonstrate that all polymers except [Cr(Bz)]$_{\infty}$ are 
metallic.
[Sc(Bz)]$_{\infty}$ is PM.
[Ti(Bz)]$_{\infty}$ is a magnetic system where the FM state is
almost isoenergetic to the most stable AFM state.  
[V(Bz)]$_{\infty}$ is quasi-HM FM and the groud state of
[Mn(Bz)]$_{\infty}$ is HM FM.
In addition, [V(Bz)]$_{\infty}$ could be
tuned to become HM FM through stretching.
The hexagonal bundles composed by [V(Bz)]$_{\infty}$ and
[Mn(Bz)]$_{\infty}$ polymers are studied.
The intrachain exchange coupling is very large due to the DE mechanism. 
We find that the interchain coupling is large enough to stabilize 
the intrachain ferromagnetic order at finite temperature.
The 1D [TM(Bz)]$_{\infty}$ polymers, especially
[V(Bz)]$_{\infty}$ and [Mn(Bz)]$_{\infty}$ with such peculiar
electronic and magnetic properties
would be ideal materials for the promising spin-polarized transport.

This work is partially supported by the National Project for the
Development of Key Fundamental Sciences in China (G1999075305,
G2001CB3095), by the National Natural Science Foundation of China
(50121202, 20025309, 10474087), by the USTC-HP HPC project, by 
the EDF of USTC-SIAS, and by the SCCAS.

\newpage
\begin{table}
  \caption{Calculated lattice constant ($c$), binding energy
    ($E_{b}$), total magnetic moment per unit cell ($M$), electronic
    ground state 
    (GS), and the energy difference between the FM and AFM states
    for the [TM(Bz)]$_{\infty}$ sandwich
    polymers (TM = Sc, Ti, V, Cr, and Mn).
    ($\Delta E=E$(FM)-$E$(AFM))} 
  \begin{tabular}{cccccc}
    \hline
    \hline
    TM& $c$(\AA)  &$E_{b}$(eV/TM)& $M$($\mu_{B}$)&GS & $\Delta E$(eV/TM)\\ 
    \hline
    Sc & 3.78       & 4.844 & 0.00           & PM  metal        &/\\
    Ti & 3.58       & 5.317 & 0.00           & AFM metal        &  0.006 \\
    V  & 3.37       & 5.334 & 0.80           & FM  metal        & -0.084 \\
    Cr & 3.30       & 2.181 & 0.00           & NM  insulator    &/\\
    Mn & 3.37       & 1.714 & 1.00           & FM  metal        & -0.250 \\
    \hline
    \hline
  \end{tabular}
  \label{table1}
\end{table}

\end{document}